\documentclass[prx,twocolumn,aps,groupedaddress,longbibliography,superscriptaddress, 10pt]{revtex4-2}
\usepackage{graphicx}
\usepackage{hyperref}
\hypersetup{
      colorlinks=true,
      citecolor=blue,
      linkcolor=blue,
      urlcolor=blue}
\usepackage{amsmath}
\usepackage{amsfonts}
\usepackage{amssymb}
\usepackage{braket}
\usepackage{bm}
\usepackage{times}
\usepackage{amsmath}
\usepackage{siunitx}
\usepackage{ragged2e}
\usepackage[dvipsnames]{xcolor}
\usepackage{nicefrac} 
\usepackage{lipsum}
\usepackage{nameref}
\usepackage{hyperref}
\usepackage{textcomp} 
\usepackage{urwchancal}
\usepackage{upgreek} 
\usepackage{amsmath} 
\usepackage{siunitx}
\usepackage{booktabs}
\usepackage{longtable}
\usepackage{scalerel}
\usepackage{tikz}
\usetikzlibrary{svg.path}
\usepackage{nicefrac} 
\usepackage{upgreek}
\usepackage{algpseudocode}
\usepackage{algorithm}
\graphicspath{{figures/}}

\newcommand{\pdd}[2]{\frac{\partial^2 #1}{\partial #2^2}}

\definecolor{orcidlogocol}{HTML}{A6CE39}
\tikzset{
  orcidlogo/.pic={
    \fill[orcidlogocol] svg{M256,128c0,70.7-57.3,128-128,128C57.3,256,0,198.7,0,128C0,57.3,57.3,0,128,0C198.7,0,256,57.3,256,128z};
    \fill[white] svg{M86.3,186.2H70.9V79.1h15.4v48.4V186.2z}
                 svg{M108.9,79.1h41.6c39.6,0,57,28.3,57,53.6c0,27.5-21.5,53.6-56.8,53.6h-41.8V79.1z M124.3,172.4h24.5c34.9,0,42.9-26.5,42.9-39.7c0-21.5-13.7-39.7-43.7-39.7h-23.7V172.4z}
                 svg{M88.7,56.8c0,5.5-4.5,10.1-10.1,10.1c-5.6,0-10.1-4.6-10.1-10.1c0-5.6,4.5-10.1,10.1-10.1C84.2,46.7,88.7,51.3,88.7,56.8z};
  }
}

\newcommand\orcidicon[1]{\href{https://orcid.org/#1}{\mbox{\scalerel*{
\begin{tikzpicture}[yscale=-1,transform shape]
\pic{orcidlogo};
\end{tikzpicture}
}{|}}}}

\def\range #1 #2 #3 {\SIrange{#1}{#2}{#3}\xspace}
\DeclareSIUnit\gauss{G}

\begin{document}

\title{Observation of Fano-suppression in scattering resonances of bosonic erbium atoms}

\author{L. Lafforgue\,\orcidicon{0009-0000-8234-6328}}
\affiliation{Universit\"at Innsbruck, Institut f\"{u}r Experimentalphysik, Technikerstr. 25, 6020 Innsbruck, Austria}
\author{N. P. Mehta\,\orcidicon{0000-0001-5450-0973}}
\affiliation{Department of Physics and Astronomy, Trinity University, San Antonio, Texas 78212, USA}
\affiliation{Institute for Theoretical Physics, Institute of Physics, University of Amsterdam, Science Park 904, 1098 XH Amsterdam, the Netherlands}
\author{J.J.A. Houwman \orcidicon{0009-0003-3342-2322}}
\affiliation{Universit\"at Innsbruck, Institut f\"{u}r Experimentalphysik, Technikerstr. 25, 6020 Innsbruck, Austria}
\affiliation{Institut f\"{u}r Quantenoptik und Quanteninformation, \"Osterreichische Akademie der \\ Wissenschaften, Technikerstr. 21A, 6020 Innsbruck, Austria}
\author{F. Claude}
\affiliation{Universit\"at Innsbruck, Institut f\"{u}r Experimentalphysik, Technikerstr. 25, 6020 Innsbruck, Austria}
\author{S. T. Rittenhouse\,\orcidicon{0000-0002-1687-7320}}
\affiliation{Department of Physics, the United States Naval Academy, Annapolis, Maryland 21402, USA}
\affiliation{Institute for Theoretical Physics, Institute of Physics, University of Amsterdam, Science Park 904, 1098 XH Amsterdam, the Netherlands}
\author{F. Ferlaino\,\orcidicon{0000-0002-3020-6291}}
\author{M. J. Mark\,\orcidicon{0000-0001-8157-4716}}
\affiliation{Universit\"at Innsbruck, Institut f\"{u}r Experimentalphysik, Technikerstr. 25, 6020 Innsbruck, Austria}
\affiliation{Institut f\"{u}r Quantenoptik und Quanteninformation, \"Osterreichische Akademie der \\ Wissenschaften, Technikerstr. 21A, 6020 Innsbruck, Austria}
\date{\today}

\begin{abstract}
The collisional properties of lanthanides exhibit remarkable complexity due to their many valence electrons, leading to an extraordinarily dense Feshbach spectrum showing signs of quantum chaos. Here we explore the situation of bosonic spin mixtures of erbium, adding the additional spin degree of freedom to the problem. We detect several inter- and intra-spin scattering resonances, exhibiting a peculiar asymmetric shape with a pronounced loss minimum. By developing a simplified multi-channel model we are able to recreate this characteristic behavior and to trace its origin to destructive interference between multiple pathways as predicted by Fano. We additionally observe a series of Fano-Feshbach resonances across multiple spin channels connected to the same molecular state, again confirmed by our theory. Our work opens the door for a detailed investigation to study multi-spin strongly-coupled scattering phenomena.
\end{abstract}


\maketitle

Dipolar quantum gases, made of strongly magnetic atoms, such as erbium~\cite{Aikawa2012bec,Aikawa2014rfd} and dysprosium~\cite{Lu2011sdb,Trautmann2018dqm}, have recently gained a lot of attention. They are enabling the experimental investigation of strongly-correlated exotic quantum phases~\cite{Norcia2021dia,Chomaz2022dpa}, including magnetic~\cite{Sachdev2008qma}, topological, and symmetry-broken phases~\cite{Dagotto1994cei,Dutta2015nsh,Chanda2025rpo}. For instance, their native long-range and anisotropic dipole-dipole interactions (DDI) allow a direct implementation of strong nearest-neighbor couplings in condensed matter Hamiltonians~\cite{Baier2016ebh,Su2023dqs}. Also, their large orbital angular momentum results in an extensive Zeeman spin-manifold, facilitating quantum simulations of large-spin Hamiltonians~\cite{Patscheider2020cde}, resource-efficient digital quantum simulation~\cite{Sawaya2020red}, or the implementation of synthetic dimensions~\cite{Bouhiron2024raq}.

Generally speaking, the two-body DDI can be decomposed into three terms~\cite{Hensler2003dri}. Two terms conserve energy and total magnetization (i.\,e.\,total spin). These are the Ising term, which leaves individual spins unchanged, and the spin-exchange term, describing flip-flop type collisions. The third term couples initial and final states with different total spin. This term, responsible for exothermic spin-relaxation analogous to the Einstein-de Haas effect, typically leads to the loss of atoms from the trap. This loss term makes dipolar spin systems inherently unstable, presenting a challenge for studying many of the above mentioned many-body phenomena~\cite{Chomaz2022dpa}. Various strategies have been explored and implemented to mitigate this effect. These include the use of tight --quasi-two-dimensional-- optical potentials and proper dipole orientation to screen the attractive part of the DDI at short range~\cite{Buechler2007sct,deMiranda2011ctq,Frisch2015udm,Barral2023ctd}. For bosons, an interesting suppression of dipolar relaxation has been observed due to the reduced overlap of the incoming-to-outgoing $s$-to-$d$ wavefunction~\cite{Pasquiou2010cod,Lecomte2025pas}. Whereas for dipolar fermions, suppression has been observed as a direct consequence of the Fermi statistics~\cite{Burdick2015fso,Baier2018roa}.

Although typically perceived as a detrimental effect, dipolar relaxation also provides a qualitatively new scattering scenario in the ultracold regime, where non-resonant inelastic background collisions are strong. In the familiar case of Feshbach resonances with alkali atoms, the presence of a resonant bound state dominates the two-body scattering, leading to a temporary trapping of the atom pair at short range for a time much longer than in non-resonant collisions. In this limit, the inelastic cross section follows the well-known symmetric Breit–Wigner profile~\cite{Breit1936cos}, revealed as well in atom loss spectroscopy~\cite{Chin2010fri}. This situation, typical of alkali with negligible to weak spin-non-conserving interactions, contrasts sharply with the dipolar case. Here, colliding pairs can either decay directly into the continuum via background dipolar relaxation, or become resonantly trapped in a quasi-bound state. This competition naturally realizes the interference mechanism first identified by Fano~\cite{fano2005absorptionspectrumnoblegases,Fano1961eoc}, where resonant and non-resonant pathways coexist and interfere. The destructive part of the interference manifests as a Fano-suppression of losses, providing a crucial pathway to stabilize excited spin states~\cite{Jie2016sot}. Remarkably, although predicted nearly a century ago, such interference has so far only been engineered by adding light-enhanced background scattering, as in photoassociation~\cite{Junker2008poa,Deb2009fri,Li2019fei} and Floquet experiments~\cite{Guthmann2025feo}. Our results demonstrate it as a direct consequence of our intrinsic dipolar interactions.

\begin{figure}[t]
        \includegraphics[width=1\columnwidth]{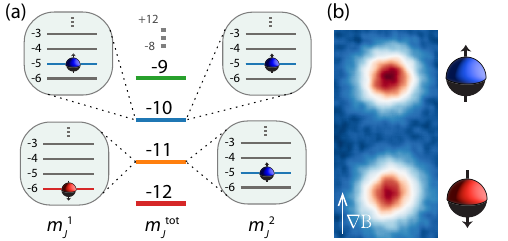}
        \caption{Spin-manifold and detection. a) Illustration of spin mixtures. Here we exemplary show a 50:50 mixture of ${m_J^1=-6}$ and ${m_J^2=-5}$ (${m_J^\text{tot}=-11}$) and a pure ${m_J^1=m_J^2=-5}$ sample (${m_J^\text{tot}=-10}$). b) Exemplary absorption image of the spin mixture with ${m_J^1=-6}$, ${m_J^2=-5}$ (${m_J^\text{tot}=-11}$).}
    \label{fig:fig1}
\end{figure}

In this work we investigate two-body scattering spin dynamics in an ultracold dipolar quantum gas, with spin encoded in one or two states of the 13-level Zeeman manifold. Through Feshbach spectroscopy across different spin combinations, we observe pronounced asymmetric loss features. Using a phenomenological  scattering model, we identify these features as a direct manifestation of the quantum interference mechanism predicted by Fano. Most notably, the destructive interference leads to a suppression of losses, enabling the preparation of metastable spin states with lifetimes exceeding hundreds of ms while retaining sizable scattering lengths. This establishes a new method for exploring spin-dependent quantum phenomena in dipolar gases. 

\begin{figure}[tb]
    \includegraphics[width=1\columnwidth]{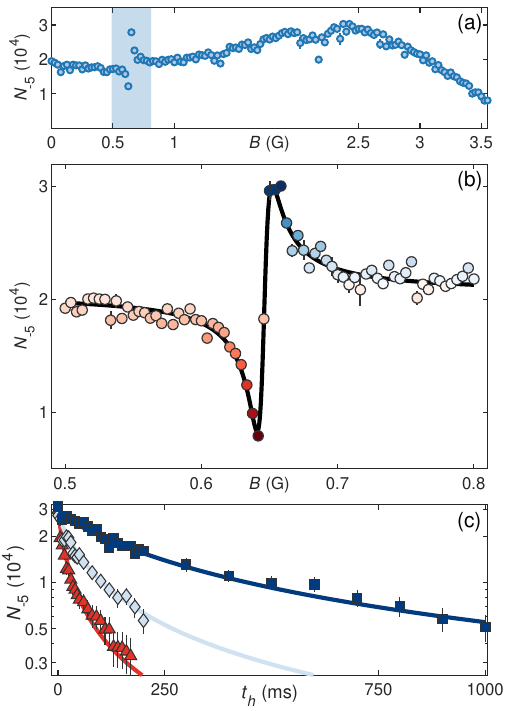}
        \caption{Observation of Fano-suppression in the $m_J^1=m_J^2=-5$ scattering channel. a) Atom number loss spectra as a function of $B$ for $m_J^1=m_J^2=-5$ ($m_J^\text{tot}=-10$) spin-polarized atoms for a hold time of $t_h=20\,$ms. b) Zoom-in on the data in a) around the Fano-shaped feature at $0.65$G. The solid line depicts a generic Fano-profile fit to the experimental data which coincides with the theory results. c) Time-resolved atom decay at three magnetic fields $B=[0.63,0.65,0.7]\,$G (triangles, squares, diamonds) across the Fano profile. The solid lines are two-body decay fits to the data, giving decay rates $L_2=[4.3(2),0.34(2),1.41(11)]\times10^{-12}\,$cm$^3$s$^{-1}$. Errorbars denote the standard error of the mean of 3-5 experimental repetitions.} 
    \label{fig:fig2}
\end{figure} 
In the experiment, we prepare an ultracold gas of about $N=3\cdot 10^4$ bosonic $^{166}$Er atoms at a temperature of $T=226(15)\,$nK just above quantum degeneracy. We confine the atoms in a harmonic trap with frequencies ($\omega_{x}$, $\omega_{y}$, $\omega_{z}$) = 2$\pi \times \left(189(4),60(10),257(4)\right)\,$\SI{}{\hertz}. All atoms are initially spin-polarized in the lowest Zeeman sublevel with $J=6$ the total electronic angular momentum and $m_J=-6$ its projection along the quantization axis~\cite{Frisch2012nlm}. The sample is prepared at a constant magnetic field of $B=1.9\,$G, collinear with gravity along the vertical ($z$) axis, maintaining spin polarization. To manipulate the spin-degree of freedom, we use a recently observed orbital clock-like transition at $1299\,$nm (linewidth $\approx 2\pi\times1\,$Hz)~\cite{Patscheider2021ooa}. This allows us to coherently transfer atoms to any desired spin state or spin combination with $>95\,\%$ efficiency in less than $100\,\mu$s by applying a sequence of Rabi pulses~\cite{claude2024omo}. For two-body scattering, the relevant quantity is the projection of the total spin given by the sum of the individual projections, $m_J^\text{tot}=m_J^1+m_J^2$, see Fig.\,\ref{fig:fig1}. For detection, we release the sample from the dipole trap and perform spin-resolved imaging by applying the standard Stern-Gerlach technique during $30\,$ms time-of-flight. From the absorption images, we extract the population of each spin state.

In a first set of experiments, each atom is prepared in the first excited spin state with ${m_J=-5}$. Dipolar relaxation occurs spontaneously as a background two-body loss mechanism, coupling $m_J^\text{tot}=-10$ to $-11$ and $-12$ and leading to atom loss at a rate $L_{2}^\text{bgr}$. Figure\,\ref{fig:fig2}a shows the dependence of the atom number $N_{-5}$ on the magnetic field $B$. We observe a broad enhancement in atom number around ${B\simeq2.5\,}$G, consistent with previous observations in chromium and dysprosium~\cite{Pasquiou2010cod,Lecomte2025pas}, where it was attributed to a reduced overlap between the incoming and outgoing scattering wavefunctions. More strikingly, we detect a narrow structure at $B\simeq0.65\,$G, highlighted in Fig.\,\ref{fig:fig2}b. The atom number exhibits an asymmetric dip–peak shape, reminiscent of an asymmetric Fano profile. Remarkably, at the point of minimum loss (destructive interference), the system is significantly longer lived than at its background, as shown in the atom number decay curves in Fig.\,\ref{fig:fig2}c.
The emergence of such a distinctive feature — never observed in alkali spinor gases — suggests that the presence of strong DDI, combined with anisotropic van der Waals dispersion forces~\cite{Frisch2014qci,Kotochigova2014cib,Maier2015eoc}, enhances coupling between scattering channels, which effectively opens various decay paths. Quantum interference between these decay paths become directly observable in the atom number as an asymmetric Fano profile. To confirm this interpretation, we model the atom number behavior around the resonance assuming a two-body rate equation 
\begin{equation}
\frac{dN}{dt}=-\frac{L_{2}}{V} N^2,
\label{eq:twobodydecay}
\end{equation}
with $V$ the effective volume~\cite{supmat} and $L_2$  the two-body decay rate following the Fano profile~\cite{fano2005absorptionspectrumnoblegases,Fano1961eoc}: 
\begin{equation}
L_{2}(B) = L_{2}^\text{bgr}+A\frac{(q\Gamma/2+B-B_0)^2}{(\Gamma/2)^2+(B-B_0)^2}
\label{eq:fanoinl2}
\end{equation}
in which $B_0$, $A$, and $\Gamma$ denote the position, amplitude and width of the resonance, respectively. $L_{2}^\text{bgr}$ is the rate coefficient away from resonance. The parameter $q$ was introduced by Fano~\cite{fano2005absorptionspectrumnoblegases} to characterize the asymmetry of the profile. We numerically solve the rate equation~\eqref{eq:twobodydecay} and fit the results to the experimental data. The model reproduces both the asymmetric loss feature versus $B$ (Fig.\,\ref{fig:fig2}b) and the time evolution (Fig.\,\ref{fig:fig2}c), confirming that the observed behavior arises from the Fano interference.

To understand the underlying microscopic processes, we develop a phenomenological scattering model. In our model, instead of the complex molecular potential of lanthanides~\cite{Kotochigova2014cib}, we take the simplified square-well model described in Ref.\,\cite{mehta2018mspr} and extend it to allow for more than one open channel. To account for the fact that at short ranges the off-diagonal elements of the DDI result in non-perturbative couplings to distant angular momentum states, we introduce three distinct couplings between scattering channels with arbitrary $\Delta m_J^\text{tot}$: entrance channel to bound state channel, entrance channel to loss channels, and bound state channel to loss channels. We can write the potential matrix of our model in the following form: 
\begin{equation}
\label{eq:mcv}
    V_{\alpha \beta}(r) = 
    \begin{cases}
    (E^{(\text{th})}_\alpha-D_\alpha)\delta_{\alpha \beta} + C_{\alpha \beta}(1-\delta_{\alpha \beta})   & \hspace{0.05in}\left(r<r_0\right)\\
    E^{(\text{th})}_\alpha \delta_{\alpha \beta}   &\hspace{0.05in}\left(r>r_0\right).
    \end{cases}         
\end{equation}
Here, $D_\alpha$ denotes the potential depth of channel $\alpha$ and $C_{\alpha \beta}$ the coupling between channels $\alpha$ and $\beta$. The model parameters are specified in units of the dipolar length $a_d = \mu_0\mu_d^2 m/8\pi\hbar^2$ and corresponding energy $E_d=\hbar^2/ma_d^2$, where $\mu_0$ is the magnetic permeability constant, $\mu_d=7\mu_B$ is the dipole moment, $\mu_B$ is the Bohr magneton, and $m$ is the atom mass. We set the width of the square well equal to the dipolar length $r_0=a_d$. The field dependence of the collision thresholds is determined by the two-atom Zeeman energy $E_\alpha^\text{th}=E_{m_J^\text{tot}}(B)= \hbar m_{J}^{\text{tot}} g_{J} \mu_B B$, where $g_J=1.1638$ is the Land\'e g-factor for the ground state~\cite{Martin1978ael}. 

We extract loss rates and atom number profiles by first solving a multichannel Schr\"odinger equation of the form $\left(\mathbf{1} H_0 +\mathbf{V}\right)\vec{\psi}=E\vec{\psi}$ with scattering boundary conditions that define the reactance matrix $\mathbf{K}$.  The solution details are given in the supplemental material~\cite{supmat}. The scattering matrix $\mathbf{S}$ is related to $\mathbf{K}$ by $\mathbf{S}=(\mathbf{1}+i\mathbf{K})(\mathbf{1}-i\mathbf{K})^{-1}$. From $\mathbf{S}$ we derive the scattering cross section to go from channel $\alpha$ to channel $\beta$, which are expressed in a basis of symmetrized spin states:
\begin{equation}
\sigma_{\beta\alpha} = \frac{\pi}{k_\alpha^2}\left|S_{\beta\alpha}-\delta_{\beta\alpha}\right|^2.
\end{equation}
This allows us to calculate $L_{2,ij}$ as the two-body loss coefficient for the rate equation (Eq.\,\eqref{eq:twobodydecay}) as
\begin{equation}
L_{2,ij} = (1+\delta_{ij})\;v_{\text{th}}\sum_{\beta\in \text{loss}}{\sigma_{\beta\alpha}(k_B T)}
\end{equation}
Here, $i$ and $j$ represent the individual spin states of the collision partners in the entrance channel $\alpha$, with the prefactor $(1+\delta_{ij})$ accounting for the fact that for $i=j$ two identical particles are lost ($\delta_{ij}=1$). 

\begin{figure}[t]
        \includegraphics[width=1\columnwidth]{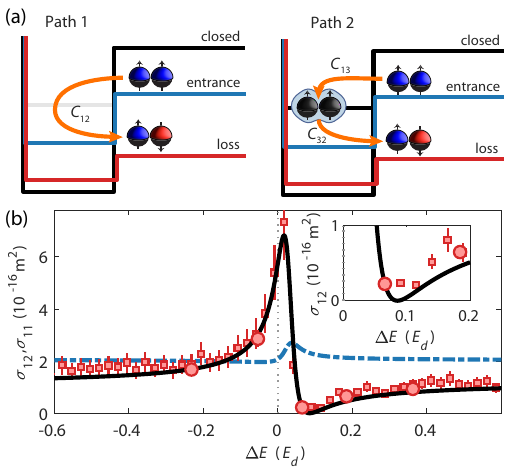}
        \caption{Square-well toy model. a) Model potentials in the case of three scattering channels (closed, entrance, loss) indicating two scattering paths: Path $1$ directly coupling the entrance and loss channel via $C_{12}$ and Path $2$ coupling the entrance and loss channel via a bound state in the closed channel with coupling strengths $C_{13}$ and $C_{32}$. b) Cross-sections $\sigma_{21}$ (inelastic, solid line) and $\sigma_{11}$ (elastic, dashed-dotted line) as a function of $\Delta E$. Markers denote the experimental $\sigma_{12}$ derived either directly from the lifetime measurements of Fig.\,\ref{fig:fig2}c (circles) or indirectly from the measured atom number of Fig.\,\ref{fig:fig2}b (squares).}
    \label{fig:fig3}
\end{figure}

\begin{figure*}[ht]
         \includegraphics[width=\textwidth]{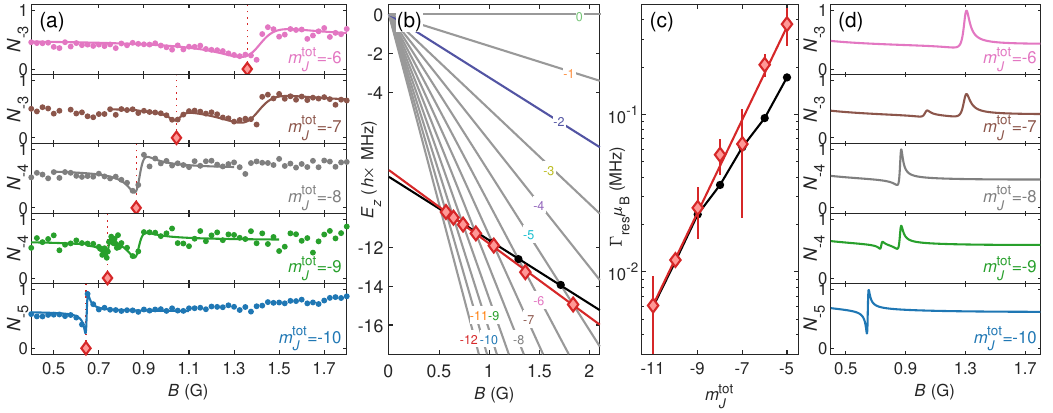}
        \caption{Molecular state crossing: (a) Loss spectra of successive spin combinations obtained in the experiment. The plots show the remaining $N_{m_J}$ after $t_h=20\,$ms from $m_J^\text{tot}=[-6, ..., -10]$  normalized to the initial atom number. Solid lines denote a fit to the experimental data. The diamonds mark the resonance position from the fits. b) Total Zeeman energy $E_{m_J^\text{tot}}$ of the scattering thresholds with the detected resonance positions for experimental (diamonds) and theory data (circles). Solid lines depict the Zeeman energy of the inferred molecular state with a magnetic moment of $\mu_\text{mol} = -2.70(4)\,\mu_B$ (experiment) and $\mu_\text{mol} = -2.328\,\mu_B$ (theory). c) Scaling of the resonance strength $\Gamma_\text{res}$ as a function of $m_J^\text{tot}$ for experimental (diamonds) and theory data (circles). The solid line shows an exponential fit through the experimental data. d) Predicted loss spectra from our toy model, see main text.}
    \label{fig:fig4}
\end{figure*}

The minimal model featuring interfering pathways consists of three channels, as shown schematically in Fig.~\ref{fig:fig3}a, similar to~\cite{Jie2016sot}. Here, the entrance ($\alpha=1$) and loss ($\alpha=2$) channels are energetically open ($E>E^{\text{th}}_\alpha$), and the closed ($\alpha=3$) channel ($E<E^{\text{th}}_\alpha$) features a bound state with binding energy $E_b$. We calculate the scattering cross sections as a function of the energy difference between the bound state and the collision threshold energy of the entrance channel together with the remaining kinetic energy $\Delta E=E_b-(E_2^{\text{th}}+k_B T)$. As shown in Fig.\,\ref{fig:fig3}b, the model predicts an asymmetric resonant behavior around $\Delta E=0$. The inelastic cross section $\sigma_{12}$ first increases substantially as $E_2^{\text{th}}$ approaches $E_b$ from below, then rapidly reaches zero and afterwards recovers to the non-resonant value. This minimum is a direct consequence of destructive interference between two scattering pathways: Atoms in the entrance channel can either decay directly to the loss channel with coupling $C_{12}$ (Path 1) or via the bound state with coupling $C_{13}$ and $C_{32}$ (Path 2). Interestingly, while the inelastic cross section can be suppressed to zero, the elastic one, related to the scattering length, remains finite for all $B$-field values. This behavior is in stark contrast to standard Feshbach resonances~\cite{Chin2010fri}. Our model allows us to quantitatively reproduce the resonance observed in Fig.~\ref{fig:fig2} as shown by the agreement with the experimentally derived $\sigma_{12}$ in Fig.~\ref{fig:fig3}b (see Ref.\,\cite{supmat} for the parameters used).

The above described physics is rather general and should equally apply to the various entrance channels of our system. To test this intuition, in a second set of experiments, we repeat atom-loss scans with different entrance channels between ${m_J^\text{tot}=-11...-5}$~\footnote{We also observe a resonance in the lowest scattering channel $m_J^\text{tot}=-12$ when probing with higher temperatures resembling a standard Breit-Rabi shape, that we interpret as resonantly increased three-body loss close to the expected position, see Ref.\,\cite{supmat}.}. The results are shown in Fig.\,\ref{fig:fig4}a. For spin mixtures we report the atom number of one of the two components. Strikingly, we observe the appearance of an asymmetric Fano profile for each of the $m_J^\text{tot}$ entrance channels studied (marked with diamonds). The resonance position ($B_0$) appears to increase quadratically with increasing $m_J^\text{tot}$.  When plotting the positions in a two-body Zeeman energy diagram, they fall along a line, see Fig.\,\ref{fig:fig4}b. This strongly suggests that a single molecular state, with constant magnetic moment, is responsible for the series of resonances by coupling to all $m_J^\text{tot}$ entrance channels. The coupling observed here between states with large differences in $m_j^\text{tot}$ is unexpected and is generally absent in alkali–alkali collisions~\cite{Chin2010fri}. In principle, dipole relaxation can only directly couple states with $\left|\Delta m_j^\text{tot}\right|\le 2$. However, within the dipole-interaction length scale, the off-diagonal components of the dipole-dipole interaction in lanthanides becomes comparable to--or even exceed--the diagonal terms, producing substantial higher-order contributions to the scattering process. Combined with the additional off-diagonal elements arising from the anisotropic dispersion interaction, this yields strong \emph{effective} coupling among all $m_j^{tot}$ states, producing the chaotic and dense Feshbach spectra observed in Dy and Er~\cite{Frisch2014qci,Kotochigova2014cib,Maier2015eoc}.

To sustain our intuition, we further analyze the observed family of Fano resonances. First, from a linear fit to the total Zeeman energy $E_{m_J^\text{tot}}$ obtained from the measured $B_0(m_J^\text{tot})$ (Fig.\,\ref{fig:fig4}b), we extract a molecular magnetic moment of $\mu_\text{mol} = -2.70(4)\mu_B$. This value is very close to the one of the $m_J^\text{tot}=-2$ scattering channel (${-2.33\,\mu_B}$), suggesting that the molecular state is attached to that scattering threshold~\cite{Frisch2015udm}. Second, we also find that the amplitude, $q$, and $\Gamma$ of the individually fitted resonances are channel-dependent. Specifically, Fig.\,\ref{fig:fig4}c shows an exponential increase of $\Gamma$ as $m_J^\text{tot}$ of the entrance channel approaches that of the molecular channel.  

In order to find out if this peculiar behavior is compatible with our theory description, we expand our model to include multiple spin channels, using the hypothesis that the molecular state is attached to $m_J^\text{tot}=-2$. We also need to include \emph{all} relevant scattering channels up to $m_J^\text{tot}=-2$. Therefore, the channel index $\alpha$ now represents the set of symmetrized spin states: 
\begin{equation}
\ket{\alpha}=\frac{\ket{m_J^1,m_J^2}+\ket{m_J^2,m_J^1}}{\sqrt{2\left(1+\delta_{m_J^1,m_J^2}\right)}}.
\end{equation}
Each collision threshold exhibits a $\left[(J^{\text{tot}}-|m_J^{\text{tot}}|)/2+1\right]$-fold degeneracy, where the brackets denote the integer part. We choose the entrance channel according to the experimental spin combinations, all other channels with equal or lower energy are considered loss channels. The channel with the highest threshold energy ($m_J^{\text{tot}}=-2$ which we choose to be non-degenerate) is engineered to support a single bound state, while we set all other channels to not support bound states, see also~\cite{supmat}. With these constraints, we first tune the model parameters to again reproduce the observed resonance in ${m_J^1=m_J^2=-5}$ collisions, and then calculate the expected scattering cross sections for all investigated spin combinations. In Fig.\,\ref{fig:fig4}d we plot the calculated atom number spectra resembling the experimental survey, showing a good qualitative agreement despite the simplicity of the model. We can again extract the resonance parameters via numerical fits and observe a remarkably encouraging agreement between theory and experiment in both resonance positions and widths. Specifically the unusual exponential scaling of $\Gamma$ is well reproduced. This justifies our assumption that the underlying interactions have an unusually strong coupling character, simultaneously connecting collision thresholds with widely different spin projections. However, further investigations are needed to clarify the underlying physical mechanisms.

In conclusion, we show that, hidden in the complexity of high-spin magnetic atoms, novel scenarios in collisional scattering physics emerge. In particular, the strong couplings that enable multiple decay pathways in the system can give rise to quantum interference phenomena in two-body scattering similar to the important quantum interference in three-body physics near an Efimov resonance~\cite{Kraemer2006efe}. 
To our knowledge, this constitutes the first direct observation of a Fano resonance manifested as a suppression of two-body losses in ultracold collisions.
These findings open new pathways for engineering interaction control and stabilizing long-range interacting spin systems in dipolar quantum gases. For instance, the suppression of losses offers an interesting opportunity to study spin-orbit quantum phenomena and the Einstein-de-Haas effect~\cite{Matsui2025oot} in bulk dipolar quantum gases. Further investigations could test the phenomenon of Fano suppression by using different single-atom spin combinations summing up to the same $m_J^\text{tot}$ state.

We thank Jeremy Hutson and Chris H. Greene for insightful discussions, and Sarah Embacher for careful reading of the manuscript. We acknowledge support from the European Research Council through the Advanced Grant DyMETEr (\href{https://doi.org/10.3030/101054500}{10.3030/101054500}), a NextGeneration EU Grant AQuSIM through the Austrian Research Promotion Agency (FFG) (No.\,FO999896041), and the Austrian Science Fund (FWF) Cluster of Excellence QuantA (\href{https://doi.org/10.55776/COE1}{10.55776/COE1}). J. J. A. H. and L. L. acknowledge funding from the Austrian Science Fund (FWF) within the DK-ALM (\href{https://doi.org/10.55776/W1259}{10.55776/W1259}). L. L.  acknowledges funding from a joint-project grant from the FWF (No.\,I-4426). 
N.~P.~M. and S.~T.~R. acknowledge support in part by the National Science Foundation through NSF PHY-2409110 and PHY-2409111, and also by the Kavli Institute for Theoretical Physics through NSF PHY-2309135.

\renewcommand{\theequation}{S\arabic{equation}}
\renewcommand{\thefigure}{S\arabic{figure}}
\setcounter{equation}{0}
\setcounter{figure}{0}

\clearpage

\section*{SUPPLEMENTAL MATERIAL}
\subsection*{Determination of the loss process}\label{appA}
\begin{figure}[h]
         \includegraphics[width=1\columnwidth]{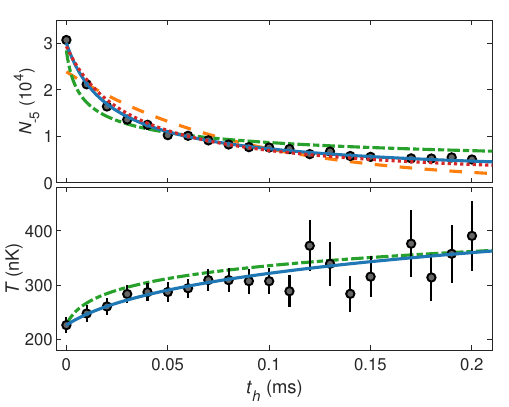}
         \caption{Analysis of the decay processes in atom number (top) and temperature (bottom). Fits with only one-body (dashed line), two-body (solid line), and three-body (dashed-dotted line) decay according to Equ.\,\eqref{equ:lossorder}. A two-body fit without temperature dependence (dotted line) is also shown for reference.}
         \label{fig:figS1}
\end{figure}

In order to determine the order of the main loss process, i.e., one-body, two-body, or three-body decay, we carefully analyze the recorded time traces around the Fano resonance of Fig.\,\ref{fig:fig2} and performed various fits, see Fig.\,\ref{fig:figS1} for an example taken at $B=0.6\,$G. Here, we take additional temperature data into account and perform a numerical fit to the following coupled differential equation~\cite{Horvath2024bec}:

\begin{equation}\label{equ:lossorder}
\begin{split}
\frac{dN(t)}{dt} & =-L_1 N(t)-L_2\frac{\beta N(t)^2}{2^{(3/2)}T(t)^{(3/2)}}-L_3\frac{\beta^2N(t)^3}{3^{(3/2)}T(t)^3},\\
\frac{dT(t)}{dt} & =L_2\frac{\beta N(t)}{2^{(7/2)}T(t)^{(1/2)}}+L_3\frac{\beta^2N(t)^2(T(t)+T_h)}{3^{(5/2)}T(t)^3}
\end{split}
\end{equation}

Here, $L_1$, $L_2$ and $L_3$ are the respective n-body loss rates, $\beta=(m\bar{\omega}^2/2\pi k_\text{B})^{(3/2)}$ and we set $T_h=0$. This allows us to determine the type of decay process. We observe that a pure two-body decay provides the best simultaneous fit to atom number as well as temperature, while three-body decay cannot reproduce short- and long-term behavior at the same time.

Nonetheless, throughout the manuscript we neglect the dependence on temperature as we do not have time-resolved temperature data of the individual spin components. A pure two-body decay fit without considering temperature still works very well to fit the atom number, see again Fig.\,\ref{fig:figS1}. 

\subsection*{Two-body recombination modeling}\label{appB}
The population dynamics of pure spin samples $m_J^1=m_J^2$ with density $n$ is governed by the differential equation,
\begin{equation}
\frac{dn_i(t)}{dt}=-L_{2,i} n_i(t)^2,
\end{equation}
where $L_{2,i}=2\sigma_{\text{loss}} v_{\text{th}}$ is the (spin-specific) two-body loss coefficient, with $\sigma_{\text{loss}}=\sum_{\beta\ne\alpha}{\sigma_{\beta\alpha}}$ the single spin-component cross section responsible for losses and ${v_{th}=\sqrt{\frac{16k_BT}{\pi m}}=7.6(3)\,}$mm/s the mean thermal relative velocity with temperature $T=226(15)\,$nK. 
For a thermal cloud, the effective volume $V=\left(\frac{4\pi k_BT}{m\bar{\omega}^2}\right)^{(3/2)}=2.35(46) \times10^{-15}$\,m$^3$ is independent of the atom number and allows us to express the density with the atom number $N_i$, resulting in
\begin{equation}
\frac{dN_i(t)}{dt}=-\frac{L_{2,i}}{V} N_i(t)^2.
\end{equation}
Finally we can integrate the differential equation to retrieve the atom number as a function of the hold time $t_h$
\begin{equation}\label{equ:loss1}
N_i(t_h)=\frac{N_i(0)}{1+L_{2,i}N_i(0) t_h/V }
\end{equation}
where $N_i(0)$ is the initial atom number. For all our spectra, we set $t_h=20\,$ms.

For spin mixtures with odd total spin, we need to describe the populations of both spin components via coupled differential equations:
\begin{equation}\label{equ:loss2}
\begin{split}
\frac{dN_i(t)}{dt}&=-\frac{L_{2,ij}}{V} N_i(t)N_j(t)-\frac{L_{2,i}}{V} N_i(t)^2.\\
\frac{dN_j(t)}{dt}&=-\frac{L_{2,ij}}{V} N_i(t)N_j(t)-\frac{L_{2,j}}{V} N_j(t)^2.
\end{split}
\end{equation}
Here, $N_i$ ($N_j$) is the atom number of spin component $m_J^1$ ($m_J^2$) with $L_{2,i}$ ($L_{2,j}$) the corresponding spin-specific two-body loss coefficient, and $L_{2,ij}$ the intra-spin two-body loss coefficient.  

\subsection*{Identifying inter- and intra-spin resonances and fitting procedures}\label{appC}
\begin{figure}[t]
         \includegraphics[width=1\columnwidth]{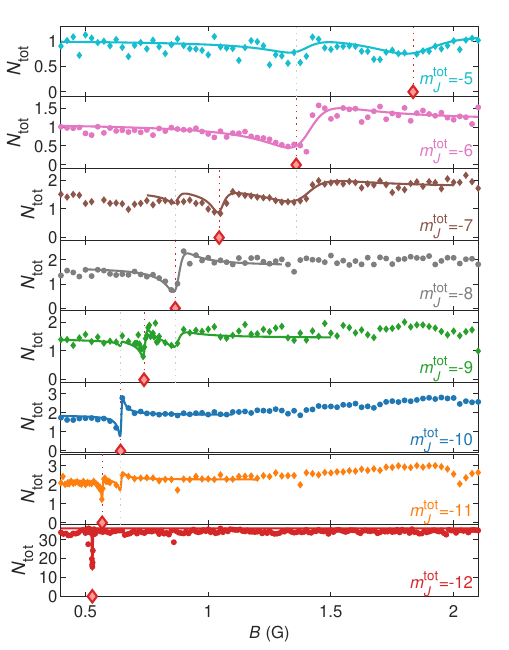}
        \caption{Spectra of successive spin combinations obtained in the experiment. The plots show the remaining atom number $N_\text{tot}$ from $m_J^\text{tot}=[-5, -11]$ as a function of magnetic field after $t_h=20\,$ms. The data for $m_J^\text{tot}=-12$ was taken at different initial conditions ($T=1.6\,\mu$K, $t_h=500\,$ms). Diamonds and the dotted vertical lines mark the positions of the corresponding resonance. Solid lines denote the fit to the data.}
    \label{fig:figS2}
\end{figure}

To identify inter- and intra-spin resonances, we first take spectra with even $m_J^\text{tot}$ consisting of singly polarized samples with $m_J^1=m_J^2=m_J^\text{tot}/2$ and identify inter-spin resonances. We then compare spectra with odd $m_J^\text{tot}$ with the pure spectra from the mixture components, see Fig.\,\ref{fig:figS2} for a complete overview. If a loss feature appears in both spin components in regions without inter-spin resonances, we identify it as an intra-spin resonance.

To fit the resonances for even total spin spectra, we first construct the loss rate coefficient $L_{2,i}(B)$, which is proportional to the loss cross section and therefore should obey a Fano shape profile:
\begin{equation}\label{equ:loss3}
L_{2,i}(B) = L_{2,i}^\text{bgr}+A_i\frac{(q_i\Gamma_i/2+B-B_i)^2}{(\Gamma_i/2)^2+(B-B_i)^2}
\end{equation}
Here, $L_{2,i}^\text{bgr}$ is a background scattering loss rate coefficient of the respective spin component, taking into account loss processes not captured by our model or caused by experimental imperfections. The Fano profile is characterized by its amplitude ${A_i,}$ the shape factor $q_i$, its strength $\Gamma_i$ and the resonance position $B_i$. We then fit the atom number by using Eq.\,(\ref{equ:loss1}) with the constructed $L_{2,i}(B)$. For Fig.\,\ref{fig:fig2}b we set $L_{2,i}^\text{bgr}$ to the minimum value derived from the independent loss measurements shown in Fig.\,\ref{fig:fig2}c. For the other even total spin spectra we take $L_{2,i}^\text{bgr}$ as an additional fitting parameter.

For all our mixed spin spectra, we first determine $L_{2,i}(B)$ for all cases with pure spin samples. For $m_J^\text{tot}=-12$ we assume $L_{2,i}(B)=0$ as it is the lowest scattering channel, and for $m_J^\text{tot}=-4$ we fit $L_{2,i}^\text{bgr}$ from independent measurements taken around $B=1.9\,$G. For each odd $m_J^\text{tot}$, we then numerically integrate the coupled equations\,(\ref{equ:loss2}) for all magnetic field values for a time $t_h=20\,$ms, taking $L_{2,i}$ and $L_{2,j}$ from the previously analyzed even spectra and using independently determined initial atom numbers as starting points. This leaves only the terms of $L_{2,ij}(B)$, again described by Eq.\,(\ref{equ:loss3}), as free fitting parameters.

For $m_J^\text{tot}=-12$ we find a $\sim$mG narrow resonance at ${B=0.528(1)\,}$G, very close to the expected value of $0.51(1)\,$G calculated from the extracted magnetic moment of the molecular state. This resonance vanishes when approaching condensation temperature, see also Ref.~\cite{Krstajic2023cot}. We interpret this as follows: The molecular state crosses the lowest scattering threshold, but as there is no two-body decay channel present, two-body scattering is purely elastic. Nonetheless, the coupling to the same bound state responsible for two-body loss features in the other collision channels can create a standard Feshbach resonance with increased three-body recombination present near the pole. The observed strong temperature dependence strengthens this interpretation, as at higher temperatures scattering processes with higher partial waves should become increasingly important, enhancing coupling and three-body loss, see Ref.~\cite{Maier2015eoc} for a more detailed discussion.

\subsection*{Multichannel square well model}\label{appD}
We can write the potential matrix of our model in the following form: 
\begin{equation}
\label{eq:mcv_supp}
    V_{\alpha\beta}(r) = 
    \begin{cases}
    (E^{(\text{th})}_\alpha-D_\alpha)\delta_{\alpha\beta} + C_{\alpha\beta}(1-\delta_{\alpha\beta})   & \hspace{0.05in}\left(r<r_0\right)\\
    E^{(\text{th})}_\alpha \delta_{\alpha\beta}   &\hspace{0.05in}\left(r>r_0\right).
    \end{cases}         
\end{equation}
Here, $D_i$ denotes the potential depth of channel $\alpha$ and $C_{\alpha\beta}$ the coupling between channels $\alpha$ and $\beta$. The model parameters are specified in units of the dipolar length $a_d = \mu_0\mu_d m/8\pi\hbar^2$ and corresponding energy $E_d=\hbar^2/ma_d^2$, where $\mu_0$ is the magnetic permeability constant, $\mu_d=7\mu_B$ is the dipole moment, $\mu_B$ is the Bohr magneton, and $m$ is the atom mass. We set the width of the square well equal to the dipolar length $r_0=a_d$. The field dependence of the collision thresholds is determined by the two-atom Zeeman energy $E^{(\text{th})}_\alpha(B)= \hbar m_{J}^{\text{tot}} g_{J} \mu_B B$, where $g_J=1.1638$ is the Land\'e g-factor for the ground state~\cite{Martin1978ael}. 

We seek solutions to the multichannel Schr{\"o}dinger equation of the form
\begin{equation}
\label{eq:mcse}
   \left[{\bf 1} H_0  + {\bf V}(r)\right]\vec{\psi}(r)=E\vec{\psi}(r).
\end{equation}
where ${\bf 1}$ is the identity matrix, and
\begin{equation}
   H_0 = \frac{\hbar^2}{2\mu}\left(-\pdd{}{r} + \frac{l(l+1)}{r^2} \right)
\end{equation}
is the kinetic energy operator in radial coordinates, with $\mu$ the reduced mass for the two colliding particles and $l$ the relative angular momentum. The  wavefunction is a vector, and its components $\psi_\alpha(r)$ represent the wavefunction in channel $\alpha$. The solution proceeds in a manner similar to that described in~\cite{mehta2018mspr}, but with the generalization that we now have more than one open channel, and therefore impose scattering boundary conditions that define the reaction matrix $\mathbf{K}$.  In the asymptotic $(r>r_0$) region, the Schr\"odinger equation is diagonal,
\begin{equation}
\label{eq:de1}
    \left(-\pdd{}{r} + \frac{l(l+1)}{r^2} - k_\alpha^2\right)\psi_\alpha(r)=0
\end{equation}
where $k_\alpha^2 = -\kappa_\alpha^2 = 2\mu(E-E_\alpha^{\text{th}})/\hbar^2$, and we define $k=\sqrt{2\mu E/\hbar^2}$. Because the present model considers $s$-waves only, we write the solution pairs as:
\begin{align}
f_\alpha(r) &= \sqrt{\frac{2\mu}{\pi k_\alpha}} \sin{(k_\alpha r)},\;\;\;f_{b,\alpha}(r) = \sinh{(\kappa_\alpha r)} \label{eq:regsols}\\
g_\alpha(r) &= -\sqrt{\frac{2\mu}{\pi k_\alpha}} \cos{(k_\alpha r)}, \;\;\;g_{b,\alpha}(r) = e^{-\kappa_\alpha r}, \label{eq:irregsols}
\end{align}
where $\{f_\alpha,g_\alpha\}$ describe solutions in open channels, and $\{f_{b,\alpha},g_{b,\alpha}\}$ describe solutions in "bound" channels that are energetically closed. Because $\mathbf{V}$ is a constant matrix in the interior region $r<r_0$, it can be diagonalized by a constant orthogonal transformation $\mathbf{\Lambda}=\mathbf{U}^T\mathbf{V}\mathbf{U}$.  Inverting the transformation renders the Schr\"odinger equation diagonal in the interior region,
\begin{equation}
(\mathbf{1}H_0+\mathbf{\Lambda})\vec{\phi}(r)=E\vec{\phi}(r),
\end{equation}
where $\vec{\phi}(r)=\mathbf{U}^T\vec{\psi}(r)$, and $\Lambda_{nn'}=\delta_{nn'}\epsilon_n$. There are as many solutions as there are channels, but each solution has only one nonzero component $[\vec{\phi}_\alpha(r)]_{n'} =\delta_{nn'}\phi_n(r)$, which is required to be regular at the origin, satisfying an equation of the form Eq.\,(\ref{eq:de1}). 
These "dressed"-state solutions are, of course, $\phi_n(r)=f_n(r)$ ($\phi_n(r)=f_{b,n}(r)$) for $E>\epsilon_n$ ($E<\epsilon_n$) given in Eq.\,(\ref{eq:regsols}) with $k_n^2=-\kappa_n^2=2\mu(E-\epsilon_n)/\hbar^2$. One can rotate back to the original basis, $\vec{\psi}=\mathbf{U}\vec{\phi}$, but in general the solutions will not match smoothly to the solutions in the exterior region $r>r_0$. A linear combination of dressed states is necessary to accomplish the matching.  We let the $\nu$th solution be
\begin{equation}
\psi_{\alpha \nu}(r)=\sum _{n} b_{n \nu } U_{n \alpha} \phi _{n}(r)
\end{equation}
and require that at $r=r_0$ both $\psi_{\alpha \nu}(r)$ and $\psi'_{\alpha \nu}(r)$ match smoothly onto the corresponding exterior solution $\psi_\alpha^{+}(r)$
\begin{equation}
\psi_{\alpha\nu}^{+}(r)=\begin{cases} 
    a_{\alpha\nu} g_{b,\alpha}(r) & \text{if } \alpha \in \text{closed} \\
    f_\alpha(r) I_{\alpha \nu}-g_\alpha(r)J_{\alpha\nu}, & \text{if } \alpha \in \text{open}.
  \end{cases}
\end{equation}

\begin{table}[t]
  \caption{\label{tbl:modelparams}}
    \begin{ruledtabular}
    \centering
    \begin{tabular}{lcc}
        Parameter & 3-channel & 31-channel \\
        \midrule
         $D_\alpha/a_{d}^2$ for $\alpha=\text{closed}$ ($m_J^{\text{tot}}=-2$) & 9.06 & $8.604$\\
         $D_\alpha/a_{d}^2$ for $\alpha=\text{open}$   & 0.967 & $1.29$ \\
         $C_{\alpha\beta}/a_{d}^2$ for ${\alpha\beta} = \{\text{entrance, closed}\}$ & 0.075 & $0.293$\\
         $C_{\alpha\beta}/a_{d}^2$ for ${\alpha\beta} = \{\text{entrance, loss}\}$   & 0.204 & $0.215$\\
         $C_{\alpha\beta}/a_{d}^2$ for ${\alpha\beta} = \{\text{closed, loss}\}$     & 0.292 & $0.423$\\
    \end{tabular}
    \label{tab:my_label}
    \end{ruledtabular}
\end{table}

Consider a case with a total of $N$ channels, $N_o$ of which are open and $N_c$ of which are closed. The matching conditions lead to a set of $2N$ equations, which by themselves are insufficient to determine the $2N+N_o$ unknown constants: $b_{\alpha\nu}, a_{\alpha\nu}, I_{\alpha\nu}$ and $J_{\alpha\nu}$.  Another $N_o$ constraints may be included by demanding that $I_{\alpha\nu}$ be equal to the $N_o\times N_o$ identity matrix, such that $\mathbf{K}=\mathbf{J}\mathbf{I}^{-1}=\mathbf{J}$. The resulting $2N+N_o$ matching equations form a linear system that uniquely determines the reactance matrix $\mathbf{K}$. The scattering matrix is related to $\mathbf{K}$ by $\mathbf{S}=(\mathbf{1}+i\mathbf{K})(\mathbf{1}-i\mathbf{K})^{-1}$, and the scattering cross section to go from channel $\alpha$ to channel $\beta$ is
\begin{equation}
\sigma_{\beta\alpha}(E) = \frac{\pi}{k_\alpha^2}\left|S_{\beta\alpha}-\delta_{\beta\alpha}\right|^2 \label{eq:sigma}
\end{equation}

Let us use indices $i$ and $j$ to represent the individual spin states of the collision partners in the entrance channel $\alpha$. As a consequence of identical particle symmetry of the scattering wavefunction, the cross section acquires an additional factor of $(1+\delta_{ij})$. The event rate coefficient, which we denote $K_{2,ij}$, determines the rate at which \emph{pairs} of atoms collide, and is given by $K_{2,ij}=\sum_\beta{\braket{\sigma_{\beta\alpha}v}}$, where $v$ is the relative velocity of the collision partners, and the brackets denote a thermal average over the Maxwell-Boltzmann distribution. 

The number of pairs in an atomic gas, however, depends on the initial spin state of the atoms. The pair density in a gas of identical atoms of type $i$ is $N_i(N_i-1)/2V\approx n_i^2 V/2$, while the pair density of distinguishable atoms of type $i$ and $j$ is $N_i N_j/V = n_i n_j V$. Thus, the atom loss rate is effectively equal to the event rate divided by a factor of $(1+\delta_{ij})$. The factor of two that arises in the cross section from identical particle symmetry effectively cancels with the factor of one-half that arises from counting pairs in a gas of identical atoms. Therefore, since each collision event results in the loss of $(1+\delta_{ij})$ atoms, the atom loss rate coefficient $L_{2,ij}$ is expressed in terms of the cross section Eq.~(\ref{eq:sigma}) as
\begin{equation}
\label{eq:losscoef}
L_{2,ij} = (1+\delta_{ij})\;v_{\text{th}}\sum_{\substack{\beta\in\text{open}\\\beta\neq\alpha}}{\sigma_{\beta\alpha}(k_B T)}
\end{equation}
For the processes considered here, we find it convenient to use the simplified expression: $\braket{\sigma_{\beta\alpha} v}\approx \sigma_{\beta\alpha}(k_B T)v_{\text{th}}$, where $v_{\text{th}}$ is the thermal relative velocity.

\begin{figure}[t]
         \includegraphics[width=1\columnwidth]{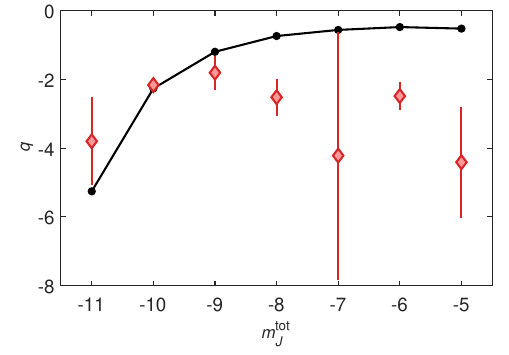}
        \caption{Evolution of the Fano shape parameter $q$. The plot shows the fitted values of $q$ as a function of $m_J^\text{tot}$ for theory (circles) and experiment (diamonds).}
    \label{fig:figS3}
\end{figure}

Table\,\ref{tbl:modelparams} displays the model parameters for both the three-channel model and the full 31-channel theory. For both models, the parameters are tuned to reproduce the resonance shown in Fig.\,\ref{fig:fig2} and Fig.\,\ref{fig:fig3}. The sequence of resonances shown in Fig.\,\ref{fig:fig4} are then predictions of the 31-channel model, where the $m_J^{tot}=-10$ resonance is in perfect agreement by construction. The uncertainty in the tuning of the model parameters is difficult to determine, since the model parameters are tuned to reproduce the number profile of Fig.~\,2 by eye. This reproduces the width and resonance position for the $m_J^{\text{tot}}=-10$ resonance to within the statistical uncertainty of the observation, as seen in Table~\ref{tbl:resonances}.  The theoretical uncertainty given in Table~\ref{tb2:resonances} is merely the statistical error in fitting the Fano profile to computed data.

\subsection*{Resonance details}\label{appE}

Our survey reveals a set of resonances connected to a single molecular channel. Table\,\ref{tbl:resonances} and Table\,\ref{tb2:resonances} denote our fitting results to the resonances depending on $m_j^\text{tot}$ to the experimental atom number profiles and the calculated scattering cross-section profiles derived from our model. Figure\,\ref{fig:figS3} plots a comparison of the derived values of the shape parameter $q$ as a function of $m_J^{\text{tot}}$.

\begin{table}[h]
  \caption{\label{tbl:resonances}}
    \begin{ruledtabular}
    \centering
    \begin{tabular}{lcccccc}
        & \multicolumn{3}{c}{Experiment} \\
        $m_j^\text{tot}$ & $B_i$ (G) & $\Gamma_i\mu_\text{B}$ (MHz) & q \\
        \midrule
         -5 & 1.84(3) & 0.37(10) & -4.4(1.6) \\
         -6 & 1.359(13) & 0.21(3) & -2.5(0.4) \\
         -7 & 1.045(17) & 0.07(4) & -4.2(3.6) \\
         -8 & 0.866(4) & 0.056(14) & -2.5(0.5) \\
         -9 & 0.739(3) & 0.026(9) & -1.8(0.5) \\
         -10 & 0.6429(3) & 0.0119(10) & -2.16(13) \\
         -11 & 0.5682(11) & 0.006(3) & -3.8(1.3) \\
    \end{tabular}
    \end{ruledtabular}
\end{table}

\begin{table}[h]
  \caption{\label{tb2:resonances}}
    \begin{ruledtabular}
    \centering
    \begin{tabular}{lccc}
        & \multicolumn{3}{c}{Theory} \\
        $m_j^\text{tot}$ & $B_i$ (G) & $\Gamma_i\mu_\text{B}$ (MHz) & q \\
        \midrule
         -5 & 1.71106(4) & 0.1724(1) & -0.5150(4)\\
         -6 & 1.28993(5) & 0.0950(1) & -0.4701(9)\\
         -7 & 1.0317(1) & 0.0621(2) & -0.553(3)\\
         -8 & 0.86025(9) & 0.0357(2) & -0.731(4)\\
         -9 & 0.73511(7) & 0.0232(2) & -1.190(6)\\
         -10 & 0.64259(2) & 0.01168(8) & -2.258(8)\\
         -11 & 0.568406(7) & 0.00601(2) & -5.25(2)\\
    \end{tabular}
    \end{ruledtabular}
\end{table}

\end{document}